\documentclass{article}      
\begin{document}

\begin{center}
\bf{ Quantization of the classical action and eigenvalue problem }\\
\vspace{3mm}
\rm{ M. N. Sergeenko\footnote{E-mail: msergeen@uic.edu }}\\
\vspace{2mm}
\end{center}

\begin{abstract}
The eigenvalue problem in quantum mechanics is reduced to quantization
of the classical action of the physical system. State function of the
system, $\psi_0(\phi)$, is written in the form of superposition of two
plane waves in the phase space. Quantization condition is derived from
the basic requirements of continuity and finiteness for $\psi_0(\phi)$ in
the whole region.
\end{abstract}
\vspace{5mm}


Quantization of the physical observables is a principal problem of
quantum theory. In three equivalent formulations of quantum mechanics,
the eigenvalue problem is solved differently and based on three independent
mathematical disciplines. Heisenberg-Dirac method is based on ``algebra'',
Schr\"odinger's approach is based on differential equations or ``analysis'',
and Feynman's path integral formulation of quantum mechanics is based on
geometry.
To make some practical calculations using any of these approaches,
a deep knowledge of the corresponding mathematical methods is required.

The quantum theory may be developed starting with classic theory and
with specific experiments that led to the replacement of classical
theory by the quantum theory. In this way, we need to introduce two
nonclassical ideas. First, energy levels of harmonic oscillators are
restricted to the values of $E=nh\nu$, with the result that energy
transfers to and from such oscillators take place in quanta with
$\Delta E=h\nu$. Second, only the probability of the transfer of a
quantum is determined by the physical state of the system.

Modern interest to the relationship between the classical quantities
and the observed behavior of quantum systems \cite{SaRo,Kay} can be
attributed to the continuing desire to attain a more thorough
understanding of quantum-classical correspondence and develop new
semiclassical approximations \cite{SeSe,SeQ}. Much of the attention
has been focused on the classical aspects of the Coulomb problem,
namely, the recent observation of wave packet recurrences in high-n
Rydberg atoms \cite{ExpData} that strongly suggests motion of the
electron in classical Kepler orbits. These experiments have
stimulated several theoretical attempts to identify Coulombic
coherent states that evolve in a manner most closely resembling the
classical dynamics. In particular, it was shown that the quantum
dynamical evolution of the electron in a Coulomb field can be
completely described by classical mechanics \cite{Kay}.

In this work we associate the eigenvalue problem with quantization of
the classical action. We introduce the state function (s.f.) $\psi_0(\phi)$ 
of the physical system in the phase space. Using requirements of continuity
and finiteness for $\psi_0(\phi)$ in the whole region, we obtain the
quantization condition which results in the discrete action variables,
i.e. quantization of the classical integrals of motion.

The most distinguished feature of particle motion in microscopic scale
is its wave properties known as the particle-wave duality. All wave
motions are periodic. Very often we are
interested not so much in the details of the orbits as in the
frequencies (or eigenvalues) of the motion. A very powerful method of
handling such systems is provided by a variation of the Hamilton-Jacobi
procedure known as the action-angle variables technique \cite{Gold}.

Value of these variables has long been demonstrated in celestial mechanics.
Recent years have seen something of a renaissance elsewhere in the use
and application of action-angle variables in problems involving the motions
of charged particles in electromagnetic fields. The adiabatic invariance
property of the action variables has led to many fruitful applications of
action-angle variables in plasma physics and in the design of particle
accelerators.

After the advent of Bohr' quantum theory of the atom, it was realized that
the quantum conditions could be stated most simply in terms of the
action-angle variables. The quantum conditions of Sommerfeld and Wilson
required that the motion be limited to such orbits for which the
action variables $J$ had discrete values that were integral multiplies
of $h$, the quantum of action. As Sommerfeld stated, the method of action-angle
variables then provided ``a royal road to quantization''. One had only to
solve the problem in classical mechanics using action-angle variables,
and motion could be immediately quantized by replacing the $J$'s with
integral multiples of Planck's constant $h$.

The mathematical justification for
quantum mechanics evolve from a number of fundamental assumptions.
First important assumption is that the state of a particle at time
$t$ completely describable by some normalized function $\psi$ that
is called the {\em state function} of the particle or system. Only
those s.f. that are physically admissable correspond to realizable 
physical states.

The s.f. itself is not experimentally observable. Another
physically necessary assumption is the probability of finding the
particle at time $t$ within a volume $dq$ centered about $q$, i.e.
$P(q,t)=\left |\psi(q,t)\right |^2$.

Consider the classical motion of a particle in Hamilton-Jacobi theory,
for which the Hamiltonian is a constant of motion and is identical 
with the total energy $E$. Hamilton's principal and characteristic 
functions are then related according to the equation

\begin{equation}  \label{ActEq}
S(q,p,t) = W(q,p) - Et,
\end{equation}
where $W$ is solution of the Hamilton-Jacobi equatioon,

\begin{equation}  \label{HJ1}
\frac 1{2m}\left(\frac{dW}{dq}\right)^2 + V(q) = E,
\end{equation}
and $p(q)=dW/dq$ is the generalzed momentum.

Since the characteristic function is independent of time, the surfaces 
of constant $W$ in configuration space have fixed locations. The motion 
of the surface in time is similar to the propagation of a wave front. 
The surfaces of constant $S$ may thus be considered as wave fronts 
propagating in configuration space.

The wave amplitude to be associated with the motion of the mechanical
particle can be taken in the form \cite{Gold} $\psi = A\,e^{iS/\hbar}$ 
or $\psi = B\,e^{-iS/\hbar}$, where $\hbar=h/2\pi$. Let us analyze the
spatial part of the wave amplitude. For the conservative system, we can
consider the following s.f. in configuration space,

\begin{equation}  \label{GenStFu}
\psi_0 = A\,e^{i\phi} + B\,e^{-i\phi},
\end{equation}
where $A$ and $B$ are constants and

\begin{equation}  \label{phiq}
\phi = \frac 1{\hbar}\int^q \frac{dW}{dq}dq.
\end{equation}

Expression (\ref{GenStFu}) can be treated as the general s.f. 
of the physical system. To build the physical s.f. in the whole
region we need to choose the boundary conditions for the problem.

In quantum mechanics, the s.f. must be continuous and finite
in the whole region $(-\infty,\infty)$. In case of free motion we have
$W=pq+{\rm const}$.
For interacting particles (the potential $V\neq 0$), the interval
$(-\infty,\infty)$ is divided by turning points (TP) given by $E-V=0$
on classically allowed regions and classically inaccessible regions.

In the classically allowed region (where $E\ge V$) near the turning point
$q_k$ the s.f. (\ref{GenStFu}) is

\begin{equation}  \label{psiC}
\psi_0^I(\phi)=A_ke^{i(\phi-\phi_k)}+B_ke^{-i(\phi-\phi_k)},
\end{equation}
and in the classically inaccessible region (where $E<V$) the function is

\begin{equation}   \label{psiF}
\psi_0^{II}(\phi)=C_ke^{-\phi+\phi_k} + D_ke^{\phi-\phi_k}.
\end{equation}

Functions (\ref{psiC}) and (\ref{psiF}) must satisfy the continuity
conditions, i.e. $\psi_0^{I}(\phi_k)=\psi_0^{II}(\phi_k)$ and
$d[\psi_0^{I}(\phi_k)]/dq=d[\psi_0^{II}(\phi_k)]/dq$, at $\phi=\phi_k$. 
Matching the functions (\ref{psiC}) and (\ref{psiF}) and their first 
derivatives at the TP $q_k$ gives

\begin{equation}
\left\{
\begin{array}{lc} \label{syst}
A_k + B_k = C_k + D_k,\\
iA_k - iB_k = -C_k + D_k,\\
\end{array}
\right.
\end{equation}
that yields

\begin{equation}
\left\{
\begin{array}{lc} \label{syspa}
A_k = \frac 1{\sqrt 2}\left(C_ke^{i\pi/4}
  + D_ke^{-i\pi/4}\right),\\
B_k = \frac 1{\sqrt 2}\left(C_ke^{-i\pi/4}
  + D_ke^{i\pi/4}\right).\\
\end{array}
\right.
\end{equation}
The connection formulas (\ref{syspa}) supply the continuous
transition of the function (\ref{psiC}) into (\ref{psiF}) at the TP
$q_k$.

Consider first the two-turning-point (2TP) problem. For this problem,
the whole interval ($-\infty,\infty$) is divided by the TP $q_1$ and
$q_2$ into three regions, $-\infty <q<q_1$ ($I$), $q_1\le q\le q_2$
($II$), and $q_2<q<\infty$ ($III$). The classically allowed region is
given by the interval $II$.

In the classically inaccessible regions $I$ and $III$ we choose the
exponentially decaying functions, i.e., $\psi_0^I(\phi) = D_1e^{\phi
-\phi_1}$ left from the TP $q_1$ (we put $C_k=0$ in Eq. (\ref{psiF})),
and $\psi_0^{III}(\phi) = C_2e^{-\phi +\phi_2}$ right from the TP $q_2$
(here we put $D_k=0$).

Then, in the classically allowed region $II$,
right from the TP $q_1$ we have, from (\ref{syspa}),
$A_1=(D_1/\sqrt 2)e^{-i\pi/4}$ and $B_1=(D_1/\sqrt
2)e^{i\pi/4}$, and the s.f. takes the form,

\begin{equation}  \label{cs1}
\psi_0^{II}(\phi)=\sqrt{2}D_1\cos\left(\phi -\phi_1 -\frac\pi 4\right).
\end{equation}
Left from the TP $q_2$ [here $A_2=(C_2/\sqrt 2)e^{i\pi/4}$ and
$B_2=(C_2/\sqrt 2)e^{-i\pi/4}$] the s.f. is

\begin{equation}   \label{cs2}
\psi_0^{II}(\phi)=\sqrt{2}C_2\cos\left(\phi -\phi_2 +\frac\pi 4\right).
\end{equation}
Here $\phi_1=\phi(q_1)$ and $\phi_2=\phi(q_2)$. We see that the
superposition of two plane waves (\ref{GenStFu}) in the phase space
results in the standing wave given by Eqs. (\ref{cs1}) and (\ref{cs2}).

Functions (\ref{cs1}) and (\ref{cs2}) should coincide at each point
of the interval $[q_1,q_2]$. Putting $\phi=\phi_2$ we have, from Eq.
Eqs. (\ref{cs1}) and (\ref{cs2}),

\begin{equation}  \label{cs12}
D_1\cos\left(\phi_2-\phi_1 -\frac\pi 4\right)=C_2\cos\frac\pi 4~.
\end{equation}
This equation is valid if

\begin{equation}  \label{eq}
\phi_2 - \phi_1 -\frac\pi 4 = \frac\pi 4 + \pi n ,~~
n = 0,1,2,\ldots
\end{equation}
and $D_1=(-1)^nC_2$.
Equation (\ref{eq}) is condition of the existence of continuous
finite s.f. in the whole region. This condition being, at
the same time, quantization condition. Taking into account (\ref{phiq}),
we have, from Eq. (\ref{eq}),

\begin{equation}  \label{qc2}
\int_{q_1}^{q_2}\frac{dW}{dq}dq = \pi\hbar\left(n +\frac 12\right).
\end{equation}

The action variables are defined as a set of independent functions of
constants of motion \cite{Gold},

\begin{equation}  \label{Ji}
J_i = \oint\frac{dW}{dq_i}dq_i,
\end{equation}
where the integration is to be carried over a complete period of
libration or rotation, as the case may be for coordinate $i$. Equation 
(\ref{qc2}) gives quantization of the classical action for the 2TP 
problem (libration). The corresponding quantized action variable is

\begin{equation}  \label{Jvar}
J_i = \oint\frac{dW}{dq_i}dq_i = 2\pi\hbar\left(n +\frac 12\right).
\end{equation}
For rotation [in this case $p(q)=P_q ={\rm const}$ and $q$ is cyclic
coordinate], the action variable is 

\begin{equation}  \label{Jrot}
J_q=2\pi\hbar n.
\end{equation}

It is known, if the Hamiltonian is conserved then the Hamilton's
principal function, $S(q_i,P_i,t)$, is the generating function to new
canonical coordinates that are all cyclic and all the momenta
are constants of integration (motion). The Hamilton-Jacobi equation
constitutes the a partial differential equation for the generating function.

In new canonical coordinates the Hamilton's characteristic function,
$W_i$, for coordinate $i$ takes the form $W_i=P_iq_i+{\rm const}$. The
constants of integration $P_i$ can be evaluated in terms of specific
initial conditions of the problem; in our case these are requirements of
continuity and finiteness for the s.f. $\psi_0(\phi)$ that results
in quantization of the action variables $J_i$ and, therefore, constants of
motion $P_i$.

Combining the above results we can write the physical s.f. in the whole 
region as

\begin{equation}
\psi_0[\phi(q)] = C_n\left\{
\begin{array}{lc}  \label{osol}
\frac 1{\sqrt 2}e^{\phi(q) -\phi_1}, & q<q_1,\\
\cos[\phi(q) -\phi_1 -\frac\pi 4], & q_1\le q\le q_2,\\
\frac{(-1)^n}{\sqrt 2}e^{-\phi(q) +\phi_2}, & q>q_2,
\end{array}
\right.
\end{equation}
where $\phi(q)=P_n q/\hbar$. The normalization coefficient, 
$C_n=\{2P_n/[\pi(n+\frac 12)+1]\hbar\}^{1/2}$, is calculated from the 
normalization condition $\int_{-\infty}^\infty \left|\psi_0(q) 
\right|^2dq=1$.

The s.f. (\ref{GenStFu}) is general for all types of
problems and allows to solve multi-turning-point problems ($M$TP,
$M>2$), i.e. a class of the ``insoluble'' problems, which cannot be
solved by standard methods. In the complex plane, the 2TP problem has
one cut between turning points $q_1$ and $q_2$, and the phase-space
integral (\ref{qc2}) can be written as the contour integral about the
cut. The $M$TP problems contain (in general case) bound state regions
and the potential barriers, i.e. several cuts. The corresponding
contour should enclose all cuts. The physical s.f. in the
whole region can be built similarly to the 2TP problem with the help
of the same connection formulas (\ref{syspa}).

Consider the $M$TP problem with $\nu$ cuts [$E-V(q)>0$ on each cut],
where all cuts are finite intervals and the effective potential
$V(q)$ (with non-communicating potential wells) is infinite between
the intervals. (In case if the potential is finite in the
whole region, we need to take into account the effect of tunneling
and the quantization condition will be more complicate \cite{SeSu}).
Then the integral around the contour $C$ can be written as sum of
contour integrals around each of the cut. Hence the $\mu=2\nu$ TP
quantization condition (and quantized action variable) can be written
as \cite{SeSe,SeQ}

\begin{equation}   \label{genCk}
J_i=\oint\frac{dW}{dq_i}dq_i = 2\pi\hbar\left(N+\frac\mu 4\right),
\end{equation}
where $N=\sum_{k=1}^{\nu}n_k$ is the total number of zeroes of the
s.f. on the $\nu$ cuts and $\mu=2\nu$ is the number of turning points 
(or Maslov's index \cite{MaslFed}, i.e. number of reflections of the 
s.f. on the walls of the potential).

To exhibit the properties of the method, let us consider the Coulomb
problem [$V(r)=-\alpha/r$] and confine our discussion to the bound case.
In spherical polar coordinates, the Hamilton-Jacobi equation,

\begin{equation}  \label{HJ}
\left(\frac{\partial W_r}{\partial r}\right)^2 + \frac
1{r^2}\left(\frac{ \partial W_\theta}{\partial\theta }\right)^2 +
\frac 1{r^2\sin^2\theta}\left( \frac{\partial W_\varphi}{\partial\varphi}
\right)^2 = 2m[E-V(r)]. 
\end{equation}
has been demonstrated to be completely separable. The motion
in each of the coordinates will be periodic - libration in $r$ and
$\theta$, and rotation in $\varphi$. The action variables are

\begin{equation}  \label{Jphi}
J_\varphi = \oint\frac{\partial W}{\partial\varphi}d\varphi =
\oint P_\varphi d\varphi,
\end{equation}

\begin{equation}  \label{Jth}
J_\theta = \oint\frac{\partial W}{\partial\theta}d\theta =
\oint\sqrt{ P_\theta^2 - \frac{P_\varphi^2}{\sin^2\theta}}d\theta,
\end{equation}

\begin{equation}  \label{Jr}
J_r = \oint\frac{\partial W}{\partial r}dr =
\oint\sqrt{2m[E-V(r)] - \frac{P_\theta^2}{r^2}}dr,
\end{equation}
where $P_\theta^2$, $P_\varphi^2$ are the constants of separation
and, at the same time, integrals of motion.

The first integral is $J_\varphi = 2\pi P_\varphi$. To calculate the integral
(\ref{Jth}) we use the method of stereographic projection. This means
that, instead of integration about a contour $C$ enclosing the classical
turning points, we exclude the singularities outside the contour $C$, i.e.,
at $\theta = 0$ and $\infty $. Excluding these infinities we have, for
the integral (\ref{Jth}), $J_\theta = I_0 + I_{\infty}$. Integral $I_0 =
-2\pi P_\varphi$, and $I_{\infty}$ is calculated with the help of the
replacement $z=e^{i\theta}$ that gives $I_{\infty} = 2\pi P_\theta$.
Therefore, $J_\theta = 2\pi(P_\theta - P_\varphi)$.

For the Coulomb potential, the integral (\ref{Jr}) can be calculated
analogously. Using the method of stereographic projection, we should
exclude the singularities outside the contour enclosing the classical
turning points $r_1$ and $r_2$, i.e. at $r=0$ and $\infty$. Excluding
these infinities we have, for the integral (\ref{Jr}),
$J_r = I_0 + I_{\infty}$, where $I_0 = -2\pi P_\theta$ and
$I_{\infty} = 2\pi i\alpha m/ \sqrt{2mE}$. The total integral is
$J_r = -(J_\theta + J_\varphi) + \pi\alpha\sqrt{-2m/E}$ that supplies the
functional dependence of $E$ upon the action variables,

\begin{equation}  \label{EJ}
E = -\frac{2\pi^2\alpha^2 m}{(J_r +J_\theta +J_\varphi)^2}.
\end{equation}
Discrete values of the action variables are given by Eq. (\ref{Jvar})
that results in the exact energy spectrum for the Coulomb problem,

\begin{equation}  \label{Ecou}
E_n = -\frac{\alpha^2m}{2(n_r+l+1)^2\hbar^2},
\end{equation}
where $l=n_\theta +n_\varphi$.
The exact eigenvalues for other central potential can be reproduced
analogously.

In conclusion, we have reduced the eigenvalue problem in quantum
mechanics to quantization of the classical action. We have considered
the classical problem in Hamilton-Jacobi theory using action-angle
variables.
We have introduced the s.f. of the system, $\psi_0(\phi)$, in
the form of superposition of two plane waves in the phase space.
Using general requirements of continuity and finiteness for $\psi_0(\phi)$
in the whole region, we have obtained the quantization condition for
the action variables and the physical s.f. for the problem.

The method has allowed us to reproduce the exact eigenvalues for known
solvable potentials and multi-turning-point problems, i.e. a class of
``insoluble'' problems, which cannot be solved by standard methods
\cite{SeClS}. It is applicable not only to separable, but non-separable
potentials, as well. For the non-separable potentials, the quantization
condition is multidimesional integral with a single quantum number for
non-separable variables.

The physical s.f. (\ref{osol}) corresponds to the main term of the 
asymptotic series in the theory of the second-order differential 
equations. For the conservative systems, we have used canonical 
coordinates that are all cyclic and all the momenta are constants of 
motion. This means that particles in stationary states move like free 
particles-waves in enclosures. This has allowed us to write the 
oscillating part of the s.f. in the form of a standing wave, which 
describes free finite motion of particles-waves in enclosures.

There is a simple connection of the method considered here with the 
Feynman path integrals \cite{Feyn}: this approach corresponds to the 
path of ``minimum'' action. The physical s.f. obtained above corresponds 
to the classical path.

{\it Acknowledgements}.
The author thanks Prof. U. P. Sukhatme for kind invitation to visit
the University of Illinois at Chicago and Prof. A. A. Bogush for
support and constant interest to this work.\\ This work was supported
in part by the Belarusian Fund for Fundamental Research.



\end{document}